\documentclass[10pt,conference]{IEEEtran}
\IEEEoverridecommandlockouts
\usepackage{cite}
\usepackage{amsmath,amssymb,amsfonts}
\usepackage{algorithmic}
\usepackage{graphicx}
\usepackage{textcomp}
\usepackage{xcolor}
\usepackage{pifont}
\usepackage{multirow}
\usepackage{flushend}
\usepackage{booktabs}
\usepackage{url}

\def\BibTeX{{\rm B\kern-.05em{\sc i\kern-.025em b}\kern-.08em
    T\kern-.1667em\lower.7ex\hbox{E}\kern-.125emX}}
\begin{document}

\newcommand{\zc}[1]{\textcolor{red}{\textbf{ZC}: #1}}
\newcommand{\mulong}[1]{\textcolor{orange}{#1}}

\title{NiCro: Purely Vision-based, Non-intrusive Cross-Device and Cross-Platform GUI Testing}

\author{\IEEEauthorblockN{Mulong Xie}
\IEEEauthorblockA{\textit{CSIRO's Data61 \&} \\ 
\textit{Australian National University} \\
mulong.xie@csiro.au}
\and
\IEEEauthorblockN{Jiaming Ye}
\IEEEauthorblockA{\textit{KYUSHU University} \\
yejjmg@gmail.com}
\and
\IEEEauthorblockN{Zhenchang Xing}
\IEEEauthorblockA{\textit{CSIRO's Data61} \\
zhenchang.xing@csiro.au}
\and
\IEEEauthorblockN{Lei Ma}
\IEEEauthorblockA{\textit{The University of Tokyo} \\
ma.lei@acm.org}
}
\maketitle

\begin{abstract}
    The diversity of devices and platforms brings significant challenges for mobile application (app) testing. 
    To ensure app compatibility and smoothness of user experience across diverse devices and platforms, developers have to perform cross-device, cross-platform testing of their apps, which is laborious.
    There comes a recently increasing trend of using a record and replay approach to facilitate the testing process.
    However, the graphic user interface (GUI) of an app running on different devices and platforms differs dramatically.
    This complicates the record and replay process as the presence, appearance and layout of the GUI widgets in the recording phase and replaying phase can be inconsistent.
    Existing techniques resort to instrumenting into the underlying system to obtain the app metadata for widget identification and matching between various devices.
    But such intrusive practices are limited by the accessibility and accuracy of the metadata on different platforms.
    On the other hand, several recent works attempt to derive the GUI information by analysing the GUI image.
    Nevertheless, their performance is curbed by the applied preliminary visual approaches and the failure to consider the divergence of the same GUI displayed on different devices.
    To address the challenge, we propose a non-intrusive cross-device and cross-platform system \textsc{NiCro}.
    \textsc{NiCro} utilizes the state-of-the-art GUI widget detector to detect widgets from GUI images and then analyses a set of comprehensive information to match the widgets across diverse devices.
    At the system level, \textsc{NiCro} can interact with a virtual device farm and a robotic arm system to perform cross-device, cross-platform testing non-intrusively.
    We first evaluated \textsc{NiCro} by comparing its multi-modal widget and GUI matching approach with 4 commonly used matching techniques.
    Then, we further examined its overall performance upon 8 various devices, using it to record and replay 107 test cases of 28 popular apps and the home page to show its effectiveness. 
\end{abstract}
\section{Introduction.}
\begin{figure}[t]
    \centering
    \includegraphics[width=0.43\textwidth]{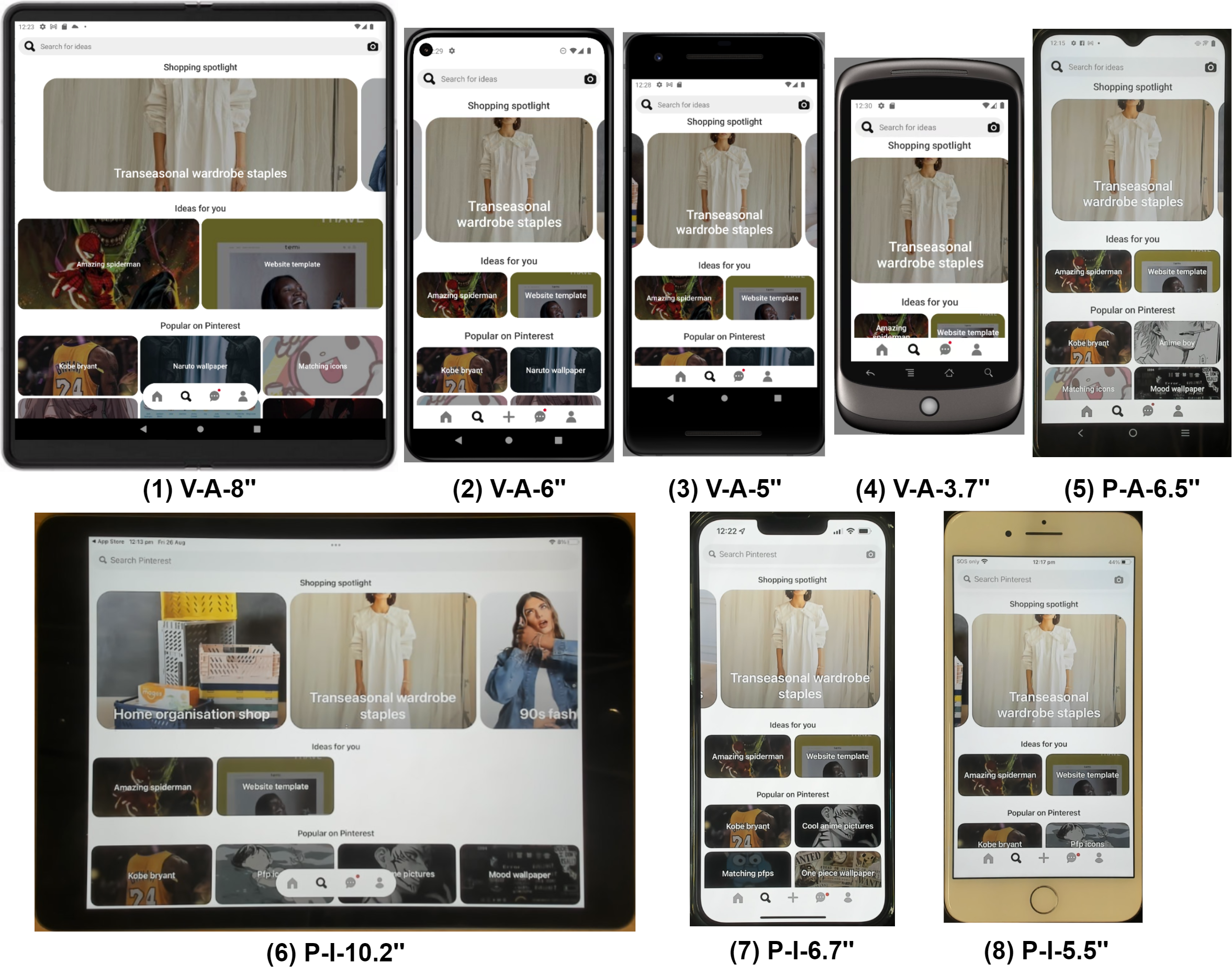}
    \caption{An app displayed on various devices. ``P`` and ``V`` indicate if the device is ``Physical`` or ``Virtual``; ``A`` and ``I`` represent ``Android`` and ``iOS`` platforms; the last number shows the screen size in inches.}
    \label{fig:devices}
    \vspace{-4mm}
\end{figure}

\begin{figure*}[t]
    \centering
    \includegraphics[width=0.85\textwidth]{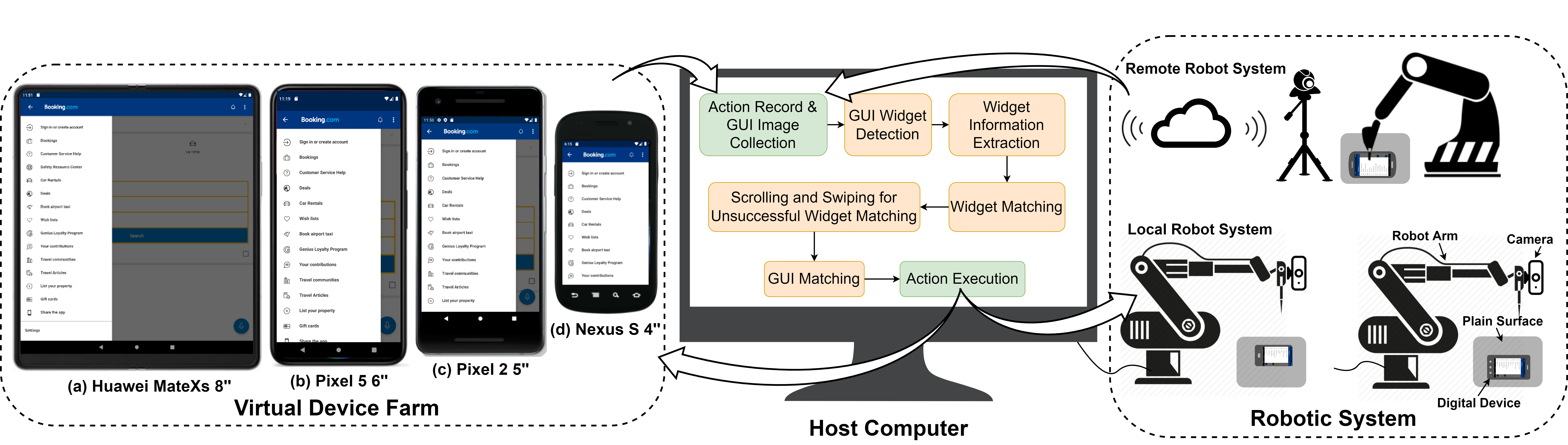}
    \caption{The architecture and high-level workflow of NiCro.}
    \label{fig:architecture}
    \vspace{-4mm}
\end{figure*}

Compatibility issue is a common critical concern of modern mobile app development, where the same mobile app could exhibit different behaviours or be rendered differently in the GUI due to the differences in a target mobile device environments (e.g., screen size, operating system) \cite{tamingandroidfragmentation, mapit}.
Fig.~\ref{fig:devices} shows some examples of an app running on different devices and environments.
The early common practice of app compatibility testing in the industry is hiring testers to execute testing procedures on diverse devices manually.
However, it involves a plethora of repetitive, laborious and time-consuming actions and pushes industries to pursue an efficient way for cross-device and cross-platform testing.

In recent years, record-and-replay testing has drawn much industrial interest, which expands automated testing with more devices and reduces the repetitive human work \cite{sikuli}.
A wide range of testing and debugging techniques (e.g., regression testing \cite{guidiff, detreduce}, failure reproduction \cite{failureproduction} and test transfer \cite{testtransfer01, testtransfer02}) are based on record and replay approaches. 
Most of these existing techniques focus on recording and replaying on the same platform, while transferring tests (recorded actions) of an app to replay on different platforms and devices has been less explored \cite{mapit}.
TestMig~\cite{testmig} makes early attempts to perform test migration from an iOS app to an Android counterpart through source code analysis.
More recent work MAPIT~\cite{mapit} proposes bi-directional UI test transfer across-platform testing for iOS and Android apps.
Nevertheless, all these approaches require intrusion into the app to acquire metadata, such as source code and run-time view hierarchy, to perform analysis.
The intrusive techniques are inapplicable or unsuitable when the underlying metadata is unavailable or not analyzable (e.g., the WebView in an app).
Moreover, they would fail to handle test cases that involve certain system operations (e.g., allow system permission) and multiple apps (e.g., share to other apps).

In contrast, non-intrusive techniques based on visual intelligence are more general, which imitate the human's way of \emph{viewing} and \emph{interacting} with GUIs. 
They only need the screenshot of the GUI (i.e., pixel image) to extract GUI information, which avoids the complexity of app intrusion involving diverse software stacks and the errors caused by the inconsistency between underlying code and the rendered GUI\cite{perceptualgrouping}.
LIRAT \cite{lirat} is the first image-driven framework proposed to solve the record and replay problem.
It obtains the GUI widget's information solely through pixel GUI images and matches widgets between different platforms based on the widget's screenshot and relative position.
However, it fails to consider the variance of GUI widget layout and placement due to the diverse screen sizes of devices (see Fig.~\ref{fig:devices}), and it does not support widget-independent operations, such as scroll.
Some works utilize a robot system for external interaction with the physical device ~\cite{robot1, robot4, roscript}.
RoScript~\cite{roscript} is the state-of-the-art non-intrusive robotic framework for record-and-replay, which recognizes touch actions as well as the target GUI widgets and drives the robot to replay.
However, RoScript only considers the tests of the same app on the same device.

Although being beneficial in many aspects by having simplicity without handling the underlying software stack, building a non-intrusive cross-device and cross-platform record and replay system faces multiple challenges.
First, different devices' hardware environments (e.g., screen size) may responsively change the GUI in terms of the layout and appearance of widgets, which relocates some widgets out of the screen and make them invisible on some small devices.
Second, an app on different platforms may experience differences in GUI design and implementation, which causes inconsistency in GUI layout and styling, further complicating the widget matching because of the change in widget's position and size.

\begin{figure*}[t]
    \centering
    \includegraphics[width=0.9\textwidth]{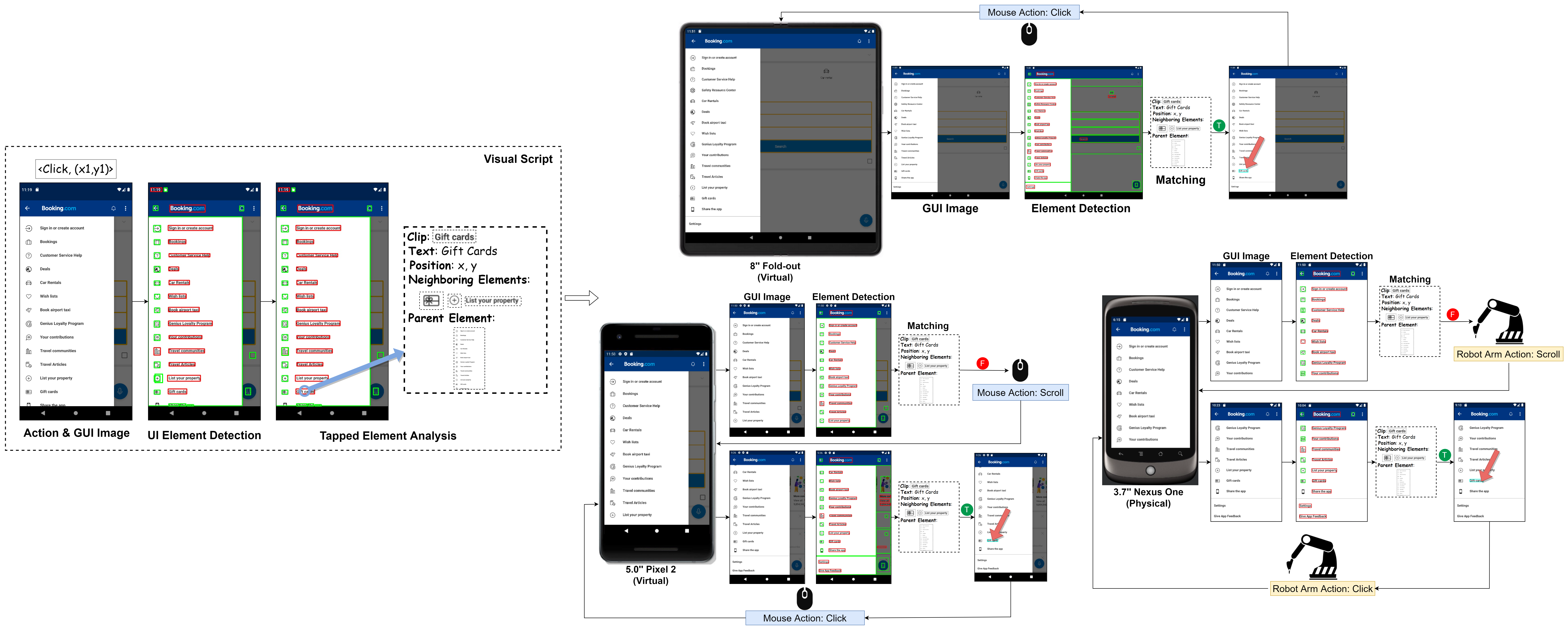}
    \caption{The summarized workflow of NiCro.}
    \label{fig:workflow}
\end{figure*}

To address these challenges and tackle record-and-replay GUI testing, we propose a purely non-intrusive cross-device and cross-platform approach NiCro.
NiCro is based on computer vision techniques to analyse GUI information from pixel GUI images without any additional need for app metadata.
Fig.~\ref{fig:workflow} summarizes the workflow of NiCro's at a high level.
It records the actions, including type and coordinates, and collects GUI images from different devices.
Next, NiCro leverages the state-of-the-art computer image-based GUI element detector UIED \cite{UIED-full-fse, xie2020uied} to detect widgets, and then extracts multi-modal attributes of each widget to represent it.
For the widget-dependent action (e.g., \textit{click}, \textit{long press} and \textit{text input}), NiCro utilizes the extracted multi-modal information to match the widget in target devices. 
For widget-independent actions (e.g., \textit{swipe horizontally} and \textit{scroll vertically}) that do not associate with any certain widgets, NiCro matches the record device's GUI with the target device's GUI to identify the ending position of the action.
At system level, NiCro comprises three key components: (1) a \textit{Device Farm}, (2) a \textit{Robotic System} and (3) a \textit{Host Computer} (see Fig. ~\ref{fig:architecture}).
The \textit{Device Farm} and the \textit{Robot System} are provided to support various virtual and physical devices for record-and-replay testing.
The \textit{Host Computer} connects and commutes with the other two components and runs NiCro's core techniques.

We performed a comprehensive evaluation from two aspects: 1) the accuracy of NiCro's multi-modal widget and GUI matching approach compared to 4 commonly used matching methods; 2) the performance of NiCro's action and test case replay with a state-of-the-art non-intrusive approach as the baseline.  
To this end, we conducted experiments upon the home page and 28 popular mobile apps spanning 14 common categories, which are installed on 8 devices, including 5 Android devices (4 emulators and 1 physical phone), 2 physical iPhones and an iPad tablet.
We randomly selected 50 widgets and 5 GUIs from each app, on which NiCro achieved 0.91 and 0.94 accuracy for widget matching and GUI matching.
We then manually created a total of 107 test cases containing 639 GUI actions that cover the general functionalities of each app. 
For each test case, we randomly select a device as the recording device (source device) and replay the actions on the rest of the other 7 devices (target devices).
Overall, NiCro achieved 0.86 and 0.85 accuracy in widget-dependent and widget-independent action replay, which accurately completes 63\% entire test cases with no correction and 89\% test cases with one-step correction, significantly outperforming its baseline.

To summarize, this paper makes the following contributions: 
\begin{itemize}
    \item A ready-to-use non-intrusive cross-device and cross-platform system\footnote{\url{https://github.com/thorxx/NiCro.git}} that can non-intrusively interact with a virtual device farm and a physical robotic system to record and replay GUI actions. 
    \item A novel computer vision-based approach to extract the multi-modal widget information and accurately match GUI widgets in the face of GUI layout and visual differences over diverse devices and platforms.
    \item A systematic evaluation of the proposed technique and a thorough analysis of the system performance to demonstrate the effectiveness of NiCro.
\end{itemize}

\section{Background}

\subsection{Cross-Device and Cross-Platform GUI Testing}
Due to the massive number and enormous diversity of modern mobile devices, a mobile app's running environments, including software and hardware, would vary dramatically and unpredictably.
This raises the strong demand for testing apps over multiple devices in the development process to tackle device fragmentation \cite{tamingandroidfragmentation}.
Although numerous works were proposed to address the compatibility issue due to device fragmentation \cite{fixfragmentation, detectfragmentation, detectfragmentation2}, most of them focus on the sole (individual) platform (e.g., Android) and are from an aspect of functionality rather than GUI.
Compared to back-end functionality, the GUI is more subject to display changes on different and heterogeneous devices, such as the screen size and resolution, because the widget's layout, appearance and even visibility may change responsively. 

As shown in Fig.~\ref{fig:devices}, the GUIs on diverse devices are differentially rendered even for the same page of the same app.
The difference exists even for devices that have the same operating platform (i.e., \textit{Device 1} to \textit{Device 5}, and \textit{Device 6} to \textit{Device 8}).
For example, many widgets on \textit{Device 1} cannot be filled in the same screen on a smaller device, such as \textit{Device 6}.
And the GUI layout may change responsively to fit the screen size.
For instance, the widgets at the bottom are arranged in rows of four on \textit{Device 5}, while they are squeezed into three-widget-row on \textit{Device 1} and two-widget-row on other smaller devices.
These variances raise more challenges in locating and matching the widget cross-devices.
Moreover, the app running on different platforms usually has to adjust its implementation, which causes some distinctions of the GUIs, hence complicating the cross-platform GUI testing. 
For instance, although \textit{Device 2} and \textit{Device 8} share a similar screen size, their GUIs are not identical (e.g., the icons at the bottom bar).
This discrepancy usually lies in the different GUI layouts and nuances of widget appearance, though the related functionalities are usually consistent.
All of the features complicate cross-device and cross-platform GUI testing.

\subsection{Non-Intrusive GUI Testing}
Non-intrusive technique, contrary to intrusive technique, does not hack into the underlying system or app to acquire data or perform action \cite{roscript}. 
It does not require connection to the target system through certain APIs and hence is not as sensitive to the variants of the programming running environment as the intrusive approaches.
This feature is desirable to testing apps running on a volatile platform, such as Android which has dozens of major versions and distributions, where the API is frequently updated or changed and requires the intrusive approach to be revised accordingly\cite{androidfragmentvisualized}.
In addition,
intrusive techniques cannot correctly acquire information from apps in many situations, such as a closed system where the app metadata is inaccessible or unanalyzable \cite{robot1}. Even for open source systems, some app metadata is still unable or inaccurate to obtain, e.g., runtime view hierarchy and metadata.

In contrast, the non-intrusive approach is more generic and less demanding.
It obtains information and interacts with the app purely through an external interface, which is analogous to human-app interaction in that we mainly use eyes to view the GUI and fingers to operate the device physically. 
The non-intrusive approaches resort to the computer vision process to ``see'' and ``understand'' GUIs \cite{whiteyolo, lirat, xie2020uied}, and some works apply robot arms to conduct the external interactions with devices, similar to human fingers \cite{robot3, robot2, roscript}.
For the non-intrusive approach,  the capacity of visual intelligence to detect and analyze the widgets on the GUI image decisively affects the overall performance.
However, most of the existing non-intrusive testing approaches rely on preliminary computer vision techniques that cannot accurately locate and recognize GUI widgets from the GUI image\cite{UIED-full-fse}, let alone the much more complicated cross-device and cross-platform testing.

\section{NiCro System Overview}
\label{sec:overview}

\subsection{System Architecture}
\label{sec:architecture}
Fig. \ref{fig:architecture} summarizes NiCro's overall architecture. 
It comprises three modules at the system level: (1) a virtual \textit{Device Farm}; (2) a \textit{Robotic System}, and (3) a \textit{Host Computer}.
The \emph{first two modules} provide various devices that NiCro interacts with in a black-box manner, and the \emph{third module} runs NiCro's core techniques to collect and process GUI information.

\subsubsection{Device Farm}
The device farm usually refers to a testing environment where a variety of devices can be accessed and controlled to test apps \cite{devicefarm1}.
A virtual device farm replaces the physical devices with app emulators. 
These emulators simulate real devices of certain models and provide the identical environment as their physical version in terms of screen size and operating system.
The device farm's scalability and flexibility enable the addition or removal of emulators easily without much additional cost, which offers a convenient platform to test apps on diverse devices.
In this work, we set up our \textit{Device Farm} through Android Studio \cite{androidstudio} that supports a convenient device manager to create and manage diverse device emulators.
NiCro non-intrusively interacts with the device manager to take a screenshot of the app and send mouse events to execute GUI actions on virtual device emulators.

\subsubsection{Robotic System}
\begin{figure}[t]
    \centering
    \includegraphics[width=0.4\textwidth]{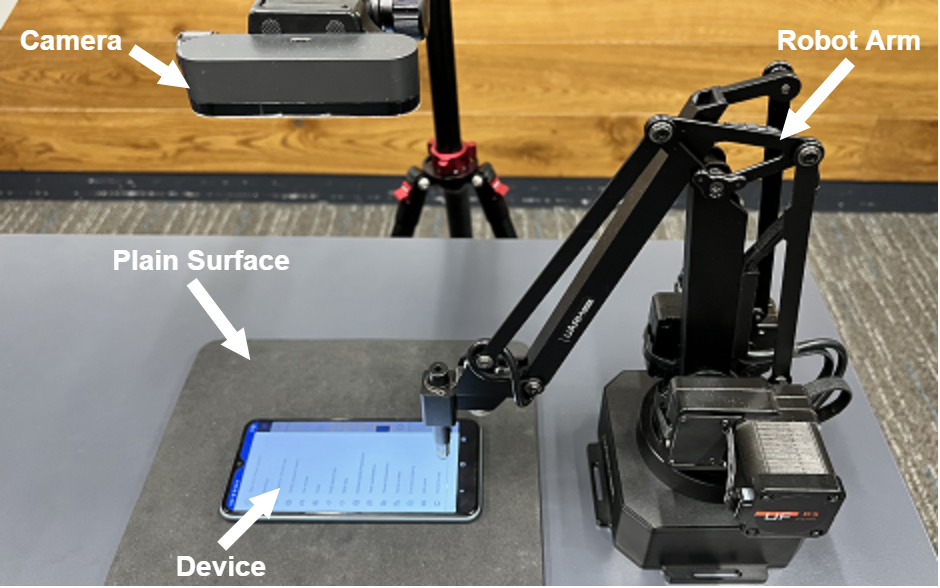}
    \caption{NiCro's Robotic System}
    \label{fig:roboticsystem}
\vspace{-4mm}
\end{figure}

NiCro supports the \textit{Robotic System} for direct and external interaction with physical devices.
Fig. \ref{fig:roboticsystem} presents the system setting, which contains three parts: a high-resolution camera, a robot arm with a stylus, and a plain surface to place the device.
The surface is simply a black pad on a desk, and the camera is set to focus on the surface to take screenshots of the devices.
NiCro is equippeded with the screen recognition technique that can extract the screen region from the picture, which is then cropped out and used as the GUI image.
The robot arm emulates a human's finger to execute the given actions on the physical device.
The robot arm and camera are highly customizable in that the hardware can be easily replaced by similar alternatives (e.g., the XY plotter used in RoScript \cite{roscript}).
The robotic system can interact with most forms of common mobile devices by simply placing the device within the surface area, which provides a convenient means for cross-device and cross-platform testing.

\subsubsection{Host Computer}
This is the central processing unit that connects the virtual devices and the robotic systems.
It collects GUI images and recorded actions from the other two components.
Specifically, it takes two kinds of input: (1) the GUI images (i.e., screenshots and screen photos) and (2) the action, including its type and coordinates recorded on a source device.
Then, it performs a series of visual analyses that consist of widget detection, multi-modal representation and matching on the GUIs of all target devices. 
Finally, it outputs the replay action, including the type and coordinate, to the \textit{Device Farm} or the \textit{Robotic System} to replay on each target device respectively.
The input from and the output to the devices only rely on the devices' external interfaces (screen).

\subsection{Visual Analytic Workflow}
\label{subsec:workflow}
The visual analytic workflow is demonstrated in Fig.~\ref{fig:architecture}'s Host Computer component and illustrated in Fig.~\ref{fig:workflow}. 

\subsubsection{Action Record and GUI Image Collection}
To start with, NiCro first records the GUI action and collects the GUI images.
An action can be performed on any one of the connected devices.
We name the device for recording action \textit{Source Device} in this work and those on which the action is replayed \textit{Target Devices}.
The vision-based design of NiCro greatly simplifies the recording process, which only needs to record the \textit{Source Device}'s GUI image (e.g., screenshot or photo) along with the action, requiring only action type and screen coordinates.
At the same time, the current GUI images on all target devices are collected for further analysis.

\subsubsection{GUI Widget Detection}
At the beginning of the visual analysis, NiCro leverages its UIED-based GUI widget detector to detect all the widgets on each GUI image.
The detection result comprises the location and class (i.e., text or non-text) for each widget on the GUI.

\subsubsection{Widget Information Extraction}
Next, NiCro analyses multi-modal attributes, in addition to the simple visual properties, from the detection result to represent the widgets for matching.
These attributes not only include the spatial and visual information, such as \textit{ Location}, \textit{Image Clip}, \textit{Text Content} and \textit{Shape}, but also contain some contextual knowledge, such as its \textit{Neighboring Widgets} and \textit{Parent Widget} if any.

\subsubsection{Widget Matching for Widget-dependent Action}
The prerequisite to replay the widget-dependent action on \textit{Target Devices} is to match the same widget with the target widget on the \textit{Source Device}.
NiCro accomplishes the matching based on the extracted multi-modal information.
Depending on the widget type (text or non-text), NiCro applies specific rules that consider the widget attributes in different priorities and finally singles out the matched widget on the \textit{Target Devices}.

\subsubsection{Expanding GUI Scope for Unsuccessful Widget Matching}
As stated in the above sections, the target widget may not appear on \textit{Target Device}'s current GUI because of the responsive adjustment or different implementation.
However, the widget is still supposed to exist to guarantee the functional consistency of a specific page in an app.
We observe that the most common reason a widget cannot be found is that the widget is allocated to the scope out of the current screen region, and it will appear when the scope is exposed to the screen region through scrolling.
Therefore, NiCro applies a screen-height scroll to the app page if no widget is matched in the current GUI and tries to match again until the target widget is identified or the margin of the page is reached.

\subsubsection{GUI Matching for Widget-independent Actions}
As for the widget-independent action, such as swiping and scrolling, NiCro replays it by matching the entire GUIs of different devices.
The swiping and scrolling start from a position on the page and move a certain distance to reach an ending position.
Due to the different display sizes of various devices, the length of the content filled on the screen would differ, and the moving distance to a certain point on the page would be inconstant and hard to transform directly. 
Therefore, NiCro turns to match the ending GUIs in different devices using its \textit{GUI Matcher} introduced in Section~\ref{sec:guimatcher}.

\subsubsection{Action Execution}
As a purely non-intrusive system, NiCro executes the actions on devices through external interactions without any intrusive control.
Specifically, it sends mouse events (e.g., click, swipe) to virtual devices in \textit{Device Farm} and controls the robot arm to tap the physical device's screen in \textit{Robotic System}, which is analogous to human-device interaction.
This simple practice covers the most common operations for mobile apps, including inputting text in which NiCro directly taps on the keyboard like the human finger.
This design greatly simplifies the action execution where \textit{Host Computer} only needs to send the action type and screen coordinates to replay the action on a \textit{Target Device}.

\section{NiCro's Visual Analytic Approach}
\label{sec:approach}

NiCro's visual analytic approach contains four major components to analyse and match GUIs.

\subsection{Widget Detector}
NiCro utilizes UIED \cite{xie2020uied}, a state-of-the-art GUI widget detector, with some revisions to detect widgets on the GUI image.
UIED is designed with the consideration of GUI's special visual characteristics.
This enables to achieve higher accuracy in GUI widget detection than deep learning models that are built for natural objects\cite{UIED-full-fse} (e.g., Faster RCNN \cite{frcnn} and YOLO \cite{yolov4}), or the methods relying on manually-crafted visual features~\cite{sift, templatematching}.
UIED contains three steps: 

\textbf{\textit{GUI Text Detection}}: The original UIED uses EAST \cite{east}, a scene text detector, to extract the text widgets.
However, the EAST is only able to locate the text area but cannot recognize the text content, while NiCro requires the text content (if any) to represent and match widgets. 
Thus we replace EAST with the widely used Google OCR tool \cite{googleocr}, which is trained to recognize a wide range of text images to detect the location and content of text widgets.

\textbf{\textit{Non-text Widgets Detection}}: UIED applies a pipeline of image processing algorithms, including gradient-based binarization, flood-filling connected component labelling \cite{ccl} and shape recognition.
The GUI-specific approach is proven to be effective \cite{UIED-full-fse}, but it suffers from the problem of losing nested widgets (e.g., widgets in a container).
NiCro tackles this problem by recognizing the container where all the nested components are kept.
In particular, NiCro identifies a widget as a container if: (1) it is rectangular and (2) none of its inside objects are connected with the widget border.
Besides, NiCro only categorizes the widgets as ``Text'' or ``Non-text''as the detailed widget category is not involved in the later process.

\textbf{\textit{Merge of Text and Non-text Widgets}}:
This step not only integrates the detection results, but also cross-checks non-text widgets and filters out false-positive non-text widgets.
UIED counts on the OCR's text results and discards non-text widgets whose bounding box intersects with any text region.

Moreover, NiCro recognizes the screen region from the GUI photo taken by the \textit{Robotic System} using heuristics: If (1) a widget's height is larger than half of the photo's height and (2) the widget contains multiple widgets and (3) the widget is not contained by any widget, it is a screen region.

\subsection{Widget Information Extractor}
\label{subsec:infoextractor}
The widget detection produces basic widget spatial and class information (i.e., size, location and text/non-text class).
However, as stated previously, the GUI and the widgets on it may change in terms of placement and appearance on different devices and platforms.
Consequently, it is necessary to extract more comprehensive multi-modal information to represent and match widgets.
Figure~\ref{fig:workflow} shows some examples of the widget information extracted by NiCro.
Overall, the extractor obtains five attributes to represent a widget.

\textbf{\textit{Spatial Location}}: The \textit{Spatial Location} is acquired from the widget detection directly and includes \textit{(Top, Left)} and \textit{(Bottom, Right)} to indicate the coordinates of the top left and bottom right corners of a widget's bounding box.

\textbf{\textit{Shape}}: This attribute implies the widget's geometric property.
It uses a 4-tuple \textit{(Width, Height, Area, Aspect Ration)} to measure the widget's bounding box, which is calculated by the \textit{Spatial Location} (i.e., (\textit{(Top, Left)}, \textit{(Bottom, Right)})).

\textbf{\textit{Image Clip}}: The extractor clips the widget's image from the GUI image as its visual information using the \textit{Spatial Location}.
Not only does the non-text widget needs the image clip to show its visual content, but also the text widget uses the image clip to present its font properties.

\textbf{\textit{Text Content}}: The \textit{Text Content} is obtained from the widget detector's OCR results directly, which can either be a single word or a line of texts.
NiCro does not combine multiple lines into a paragraph to avoid incorrect merging and tricky parameter adjustment.
So, a chunk of text that contains several lines would be identified as separate text widgets.
Note that some non-text widgets also have text content if there is a piece of text within it (e.g., the text button).

\textbf{\textit{Surrounding Widgets}}: NiCro also examines the widget from a contextual aspect.
This information is conducive to identifying the widget on the \textit{Target Device}, in particular when there are multiple widgets with the same appearance on the GUI.
A widget's parent and neighbours usually remain constant even when the GUI changes responsively to fit different devices.
Thus, the extractor includes the \textit{parent widget} (e.g., container) and the closest \textit{adjacent widgets} in four directions (i.e., up, down, left and right) for a widget.
Note that the \textit{parent} could be \textit{None}, and the number of the \textit{neighboring widgets} is 0 to 4 depending on the specific widget.

\subsection{Widget Matcher}
\label{sec:widgetmatcher}
NiCro utilizes the widget's attributes extracted in the previous step to match its counterparts in other devices. 
These attributes capture the comprehensive information of the widget and impose different degrees of influence on the matching result.
Therefore, different types (i.e., text and non-text) of widgets put distinct priorities upon these attributes while comparing.
For the sake of presentation, we denote the \textit{Source Device} GUI where NiCro records the action as $G_s$, the \textit{Target Device} GUI where the action is replayed as $G_t$.
We also denote the widgets in $G_s$ as $W_s$, and the ones in $G_t$ as $W_t$.

\subsubsection{Text Widget Matching}
\label{sec:textmatching}
The \textit{Text Content} is the most decisive factor for matching two text widgets.
However, two matched text widgets may not contain identical content.
Occasionally, a line of text in a large device's GUI would be broken down into several lines in a small device to fit the screen.
As stated in Section~\ref{subsec:infoextractor}, NiCro's widget information extractor keeps the line-level texts but does not merge multiple lines into a paragraph, so that the text line in the large GUI would be divided into several text widgets in the small GUI.
Therefore, the \textit{Text Content} matching follows the rule: If $W_s.TextContent \subseteq W_t.TextContent$ or $W_t.TextContent \subseteq W_s.TextContent$, then $W_s$ is a matched candidate to $W_t$.
If no widget in $G_t$ is matched \textit{Text Content} with $W_s$, the approach will not apply any other attributes but determines that there is no matching text widget in this GUI.

After matching by this rule, there can be multiple widgets $W_t^i (i \in \{1..n\})$ in $G_t$ that match with $W_s$.
If none of the $n$ $W_t^i$ has the same content as each other, which indicates they are just separate parts of $W_s$, \textit{Widget Matcher} selects the one with the longest content as the matched target.
Otherwise, if $W_s$ is matched with several widgets that share identical text content, the approach resorts to other attributes to filter out the candidates.
The attributes are checked from simple to complex in order, until only one widget is left.
(1) It compares the \textit{Shape} between $W_s$ and all candidates $W_t^i$.
The difference between any value in the 4-tuple \textit{(Width, Height, Area, Aspect Ration)} for the two widgets should be less than a specified threshold.
(2) For the remaining candidates, it encodes the their \textit{Image Clip}s through a ResNet50 model~\cite{resnet50} and calculates the cosine similarity~\cite{cosinedistance} between encodings of $W_t^i$ and $W_s$.
The candidate with less than 80\% similarity will be discarded.
(3) If there is still more than one candidate left, the \textit{Widget Matcher} matches the \textit{Surrounding Widgets} of $W_s$ and $W_t^i$ to single out the final target.

\subsubsection{Non-Text Widget Matching}
The non-text widgets usually contain more complicated visual features than text widgets do.
For example, the image and icon widget conveys information through pictures or symbols, and the button has a visible boxing area with colour filled or a line boundary.
Therefore, the primary attribute for the non-text widget to be matched is the \textit{Image Clip}. 
The approach first encodes the \textit{Image Clip}s through a ResNet50 model and computes the cosine similarity between encodings.
However, there can be multiple widgets with the same appearance on the GUI, which requires additional information, other than the image, to match.
So the \textit{Widget Matcher} includes the widget whose similarity is larger than 80\% as the candidate.
On the other hand, if no widget reaches the similarity threshold, the approach concludes that no widget on this GUI matches the target widget.

Similar to the text widget matching, in the case that multiple candidates are selected, the \textit{Widget Matcher} considers other attributes.
It starts with the \textit{Shape} (i.e., \textit{(Width, Height, Area, Aspect Ration)}).
But since there can be some nuances in \textit{Shape} for the same widgets on different devices, the approach allows a small degree of difference but filters out those too distinguishable.
Finally, if the \textit{Shape} still fails to single out the final widget, the approach chooses the candidate that has the most matched \textit{Surrounding Widgets} with the target widget.

\begin{table}[t]
	\centering
	\caption{Devices used in the experiment}
	\vspace{-1mm}
	\resizebox{0.40\textwidth}{!}{
	    \begin{tabular}{l c c c c }
    		\toprule
    		\textbf{\#} & \textbf{Model} & \textbf{OS} & \textbf{Size} & \textbf{Type}\\
    		\toprule
    		D1 & 8'' Fold-out & Android 12.0 & 2,200 x 2,480 & Virtual\\
    		D2 & 6.5'' Vivo Y21 & Android 12.0 & 1,600 x 720 & Physical\\
    		D3 & 6'' Pixel 5  & Android 12.0 & 1,080 x 2,340 & Virtual\\
    		D4 & 5'' Pixel 2  & Android 11.0 & 1,080 x 1,920 & Virtual\\
    		D5 & 3.7'' Nexus One & Android 11.0 & 480 x 800 & Virtual\\
    		D6 & 6.7'' iPhone 13 Pro Max & iOS 15.4 & 2,778 x 1,284 & Physical\\
    		D7 & 5.5'' iPhone 8 Plus & iOS 14.8 & 1,920 x 1,080 & Physical\\
                D8 & 10.2'' iPad 7th Gen & iOS 15.4 & 2,778 x 1,284 & Physical\\
    		\toprule
    	\end{tabular}
	}
	\vspace{-4mm}
	\label{tab:devices}
\end{table}

\begin{figure*}[t]
    \centering
    \includegraphics[width=0.93\textwidth]{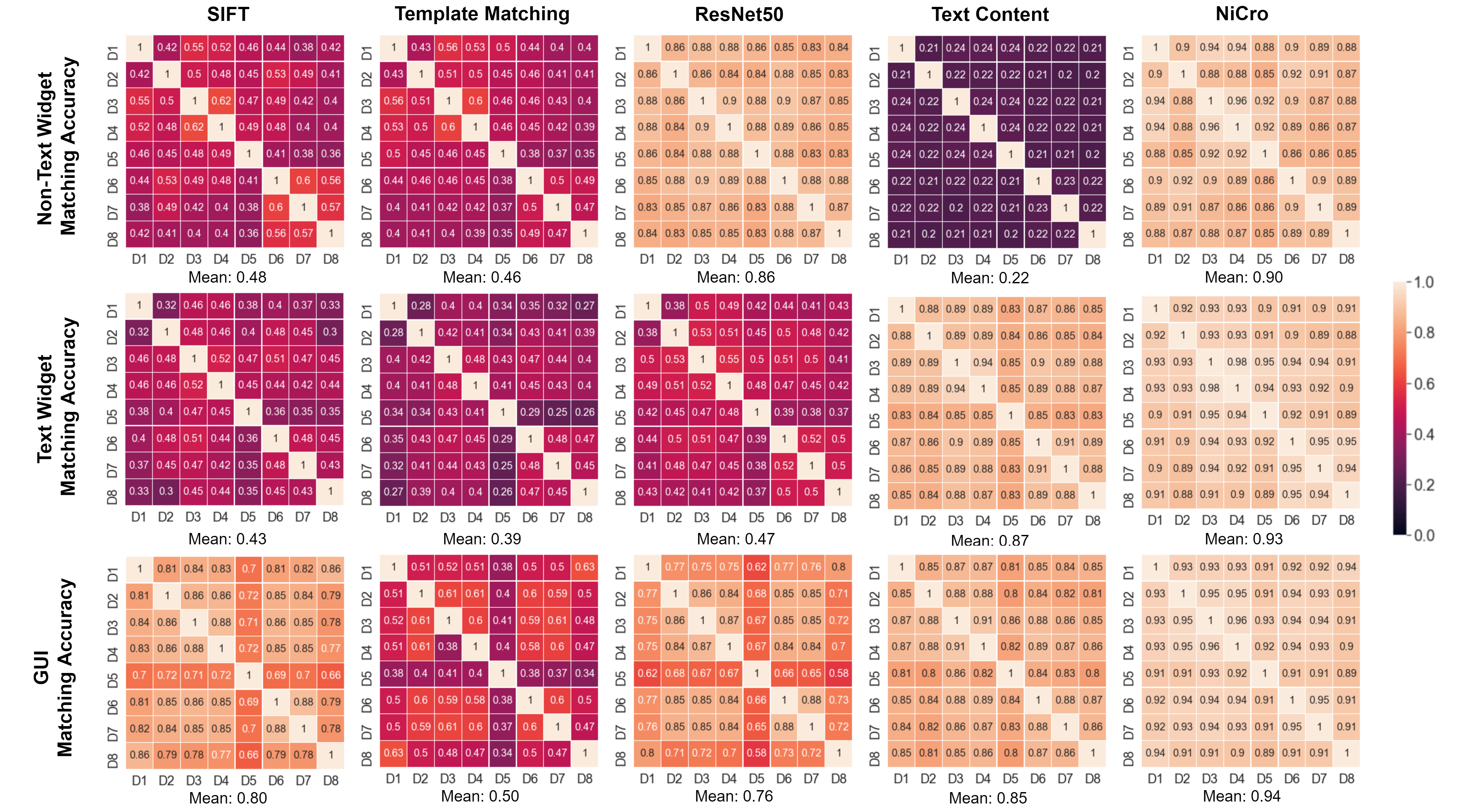}
    \caption{Experimental result of widget matching}
    \label{fig:eval_matching}
    \vspace{-4mm}
\end{figure*}

\subsection{GUI Matcher}
\label{sec:guimatcher}
NiCro has to match the entire GUIs in two situations: (1) to determine if the page reaches the margin and (2) to check if the widget-independent action (i.e., swipe horizontally and scroll vertically) is complete on a different \textit{Target Device}. 

\subsubsection{GUI Matching on Same Device}
NiCro is a purely non-intrusive system that is unable to be informed by the underlying system about the current GUI's position on the page.
Therefore, to examine if the GUI is at the margin of a page (e.g., page bottom), NiCro checks if the GUIs are identical before and after the widget-independent action.
For example, if the GUI keeps unchanged after a scroll or swipe, it means the page has no overshooting part of the screen region in the operating direction and reaches its margin.
In this case, the identical GUIs can be matched straightforwardly by comparing the pixel images.
This simple practice is effective as the content on the same page is unlikely to change in a short time during the action execution.
Even if some parts (e.g., Ads) of the GUI occasionally change, NiCro tries to execute the action and match the GUIs again until it finally finds the page has already reached its margin.

\subsubsection{GUI Matching on Different Devices}
While replaying the widget-independent action on different \textit{Target Devices}, the distance of scrolling or swiping would vary dramatically due to the different lengths of content filled on screens with diverse sizes and resolutions.
Hence, to determine whether the page on the \textit{Target Device} reaches the same or similar position as the recording \textit{Source Device}, the \textit{GUI Matcher} tries to match all the widgets on the two GUIs.
In particular, it defines the GUI from the device with a smaller screen as $G_1$ and the GUI from the larger device as $G_2$.
If all the widgets in $G_1$ are successfully matched to a counterpart in $G_2$, NiCro treats these two GUIs as matched.


\section{Evaluation}
We conduct experiments and analyses in two aspects to evaluate NiCro: (1) the performance of NiCro’s core multi-modal matching approach for widgets and GUIs and (2) its capability to complete the record and replay task for cross-device and cross-platform UI testing.

\subsection{Experiment Setup}
To investigate NiCro's cross-device capability, our evaluation leverages 5 different Android devices, including physical and virtual ones, which have a wide range of screen sizes from 3.7'' to 8.0''.
For the cross-platform replay, our evaluation also includes two iPhone devices in different models and an iPad.
Table~\ref{tab:devices} shows the details of the experimental devices and Fig.~\ref{fig:devices} demonstrates their GUIs.
Due to the current \textit{Device Farm} in NiCro does not support iOS, we only use physical iPhone devices that are interacted with the \textit{Robotic System}.

Regarding the subject mobile apps, we select a total of 28 apps spanning 14 frequently-used categories (see Table~\ref{tab:eval_actions}).
Furthermore, we record and replay actions on the device's home page.
For example, tapping the icon to open an app and swiping \& scrolling on the home page.
The app selection considers the following three perspectives: (1) it covers a range of most popular app categories to show NiCro's generalization capability; (2) the apps are widely downloaded so that they are stable to function in different devices and (3) the apps support both Android and iOS as to test NiCro's cross-platform ability.

NiCro's virtual \textit{Device Farm} is implemented based on Android Studio in version 2021.1.1.23, which provides a variety of emulators with various Android versions. 
For the \textit{Robotic System},  we leverage an \emph{UArm Swift Pro Robot Arm} manufactured by Ufactory~\cite{ufactory}, a 4-axis robot that holds a stylus pen to mimic a human finger to interact with the touch screen. 
The camera used in our system is a \emph{4K Logitech BRIO Webcam}~\cite{logitech}, and the surface is simply a black pad placed under the camera.
Finally, the \textit{Host Computer} we use in the experiment is a desktop and runs Ubuntu 18.04.

\subsection{Accuracy of Widget Matching and GUI Matching}
The evaluation first examines the matching accuracy for widgets and GUIs.
We implemented a range of image matching methods and tested them on the same experimental data with NiCro.
Different from many previous image-based non-intrusive testing systems that solely or primarily rely on the widget image~\cite{sikuli,roscript,jauto}, NiCro adopts multi-modal information to represent and match GUI widgets and the entire GUI.
The evaluation includes the techniques applied in two closest non-intrusive testing works and two individual single-modal approaches NiCro utilizes.
In particular, we evaluate (1) \textit{Template Matching} used in RoScript; (2) \textit{SIFT} used in LIRAT; (3) \textit{ResNet} that NiCro applies to encode widget image and (4) \textit{Text Content Matching} in NiCro to match text widget's content and texts on some non-text widgets.

In the experimental process, we iteratively select a \textit{Source Device} and use the rest of the other 7 devices as \textit{Target Devices}.
This yields a 8 x 8 matrix of results for each evaluated method, as shown in Fig.~\ref{fig:eval_matching}.
From the collected 28 apps and the home page, we randomly chose 25 text widgets and 25 non-text widgets from each app.
Overall, 1,450 (29 x (25 + 25)) widgets are used as the experimental subjects, and every widget is guaranteed to be visible on all of the 8 devices while being matched, although their relative position and size would vary in different devices.
Besides, we randomly select 5 GUIs from each app on the 8 devices to examine the GUI matching, which brings a total of 1,160 (5 x 29 x 8) GUIs.
Then, we apply the matching algorithms to find the most matched widget for a \textit{Target Widget} and the most matched GUIs for a \textit{Target GUI} on a \textit{Target Device}.
Finally, we manually inspect if the matching is correct and calculate the matching accuracy.

\subsubsection{Non-Text Widget Matching}
Fig.~\ref{fig:eval_matching} present the experimental results.
Overall, the \textit{Text Content} and \textit{Template Matching} perform worst in non-text widget matching and text widget matching (0.23 and 0.39), while \textit{NiCro}'s approach achieves the best performance in both cases (0.9 and 0.93).
For non-text widgets, there are 28\% of them have text content (e.g., button with text in the centre), while the rest of 72\% do not contain any text information (e.g., image, icon).
Since the \textit{Text Content} can only match widgets through text, it is reasonable that it performs poorly on the non-text widgets.
The low accuracy for \textit{SIFT} is because many of the non-text widgets, such as icons, do not have complicated textures so \textit{SIFT} fails to extract sufficient features points as the widget image's descriptor for accurate matching. 
\textit{Template Matching} simply slides the \textit{Target Widget}'s image over the \textit{Target Device}'s GUI image to match the most similar patch. 
It is highly subject to the scales of the template image (\textit{Target Widget}'s image) and the input image (\textit{Target Device}'s GUI image) \cite{templatematchingopncv}, and thus it can only perform well when the \textit{Source Device} and the \textit{Target Device} have similar screen size and the corresponding widgets' appearance are consistent.
\textit{ResNet50} also achieves a high accuracy because it, as a well-pre-trained image encoder, represents the image's feature effectively, thereby minimizing the cosine distance between the embeddings of matched widget images.

\begin{table}[t]
	\centering
	\caption{Comparison between Closest works and NiCro}
	\resizebox{0.47\textwidth}{!}{
	    \begin{tabular}{l c c c c c c}
    		\toprule
    		\textbf{Work} & \textbf{Intrusion} & \textbf{Input} & \textbf{Cross-P} & \textbf{Cross-D}  & \textbf{Wid-Indep} & \textbf{TC}\\
    		\toprule
    		\textbf{MAPIT} & Intrusive & Metadata & \checkmark & \textit{x} & \textit{x} & 27.1\%\\
    		\textbf{LIRAT} & N-Intrusive & Image & \checkmark & \textit{x} & \textit{x} & 38.3\%\\
    		\textbf{RoScript} & N-Intrusive & Image & \textit{x} & \textit{x} & \checkmark & 100\%\\
                \textbf{NiCro} & N-Intrusive & Image &\checkmark & \checkmark & \checkmark & 100\%\\
    		\toprule
    	\end{tabular}
	}
	\vspace{-4mm}
	\label{tab:comparisons}
\end{table}

\subsubsection{Text Widget Matching}
On the other hand, all the three image matching algorithms \textit{SIFT}, \textit{Template Matching} and \textit{ResNet50} fail to match precisely for text widgets for two major reasons: (1) the visual and texture features of text characters are not as distinct as non-text images~\cite{cnntext} and (2) a long line of text on a large device would be broken down to multiple short lines on a small device.
As the analysis in \textit{Text Widget Matching} (Sec~\ref{sec:textmatching}), the short lines are still supposed to be matched to the long one because they are the partial components, while their visual properties are apparently different.
However, the \textit{Text Content} reaches a high matching accuracy (0.87), which indicates the text content is decisive for distinguishing text widgets.
Nevertheless, using \textit{Text Content} solely still fails in some cases where additional information is required, especially when there are multiple same or similar text pieces on a GUI.

\subsubsection{GUI Matching}
The GUI matching accuracy for the three image matching algorithms is interestingly different from their accuracy for individual widgets.
\textit{SIFT} performs much better in GUI matching (0.80) because the entire GUI is much larger than the single widget and contains more SIFT features.
It is even better than \textit{ResNet} (0.76), as \textit{SIFT} is scale-invariant but \textit{ResNet} is more subject to the GUI size.
\textit{Template Matching} reaches a noticeably better accuracy for GUIs from similar-size devices, while it is incapable of matching GUIs from devices of various size that contains the different length of content because this method conducts a pixel-level matching.
Matching GUIs using \textit{Text Content} also produces an acceptable accuracy (0.85).
Here we adopt the same text content matching strategy as NiCro does, in which if the text content in the smaller GUI can be matched to the larger GUI, it regards them as matched.
NiCro's GUI matching approach based on multi-modal widget matching obviously outperforms the compared methods and achieves an accuracy of 0.94.

\begin{table*}[!ht]
    \centering
    \caption{Experimental results of action replay and test cases replay using NiCro \& RoScript}
    \resizebox{0.99\textwidth}{!}{
\begin{tabular}{llcccccccccccccccc}
    \toprule
        \multirow{2}{*}{\textbf{Category}} & \multirow{2}{*}{\textbf{App}} & \multirow{2}{*}{\textbf{\#TC}} & \multicolumn{2}{c}{\textbf{0-Correction Acc}} & \multicolumn{2}{c}{\textbf{1-Correction Acc}} & \multicolumn{4}{c}{\textbf{\#Widget Dependent Action}} & \multicolumn{2}{c}{\textbf{Accuracy}} & \multicolumn{3}{c}{\textbf{\#Widget Independent Action}} & \multicolumn{2}{c}{\textbf{Accuracy}} \\ \cline{4-18}
        ~ & ~ & ~ & \textbf{RoScript} & \textbf{NiCro} & \textbf{RoScript} & \textbf{NiCro} & \textbf{Click} & \textbf{L-Press} & \textbf{Input} & \textbf{Total} & \textbf{RoScript} & \textbf{NiCro} & \textbf{Swipe(H)} & \textbf{Scroll(V)} & \textbf{Total} & \textbf{RoScript} & \textbf{NiCro} \\ \toprule
        
        \multirow{2}{*}{\textbf{Travel}} & Booking & 4 & \multirow{2}{*}{1} & \multirow{2}{*}{6} & \multirow{2}{*}{2} & \multirow{2}{*}{8} & 15 & 0 & 2 & 17 & \multirow{2}{*}{0.41} & \multirow{2}{*}{0.89} & 1 & 2 & 3 & \multirow{2}{*}{0.25} & \multirow{2}{*}{0.88} \\ \cline{2-3} \cline{8-11} \cline{14-16}
        ~ & AirBNB & 4 & ~ & ~ & ~ & ~ & 18 & 0 & 1 & 19 & ~ & ~ & 2 & 3 & 5 & ~ & ~ \\ \hline
        \multirow{2}{*}{\textbf{Education}} & Coursera & 4 & \multirow{2}{*}{1} & \multirow{2}{*}{6} & \multirow{2}{*}{2} & \multirow{2}{*}{7} & 14 & 0 & 1 & 15 & \multirow{2}{*}{0.43} & \multirow{2}{*}{0.90} & 3 & 3 & 6 & \multirow{2}{*}{0.22} & \multirow{2}{*}{0.89} \\ \cline{2-3} \cline{8-11} \cline{14-16}
        ~ & Sololearn & 3 & ~ & ~ & ~ & ~ & 14 & 1 & 0 & 15 & ~ & ~ & 2 & 1 & 3 & ~ & ~ \\ \hline
        \multirow{2}{*}{\textbf{Browser}} & Chrome & 4 & \multirow{2}{*}{1} & \multirow{2}{*}{5} & \multirow{2}{*}{1} & \multirow{2}{*}{7} & 16 & 2 & 2 & 20 & \multirow{2}{*}{0.40} & \multirow{2}{*}{0.89} & 1 & 5 & 6 & \multirow{2}{*}{0.27} & \multirow{2}{*}{0.91} \\ \cline{2-3} \cline{8-11} \cline{14-16}
        ~ & Firefox & 4 & ~ & ~ & ~ & ~ & 13 & 1 & 2 & 16 & ~ & ~ & 1 & 4 & 5 & ~ & ~ \\ \hline
        \multirow{2}{*}{\textbf{Email}} & Gmail & 4 & \multirow{2}{*}{0} & \multirow{2}{*}{5} & \multirow{2}{*}{0} & \multirow{2}{*}{7} & 14 & 0 & 2 & 16 & \multirow{2}{*}{0.34} & \multirow{2}{*}{0.93} & 3 & 4 & 7 & \multirow{2}{*}{0.25} & \multirow{2}{*}{1.00} \\ \cline{2-3} \cline{8-11} \cline{14-16}
        ~ & Outlook & 3 & ~ & ~ & ~ & ~ & 10 & 0 & 1 & 11 & ~ & ~ & 2 & 3 & 5 & ~ & ~ \\ \hline
        \multirow{2}{*}{\textbf{News}} & BBC & 4 & \multirow{2}{*}{1} & \multirow{2}{*}{6} & \multirow{2}{*}{2} & \multirow{2}{*}{7} & 13 & 0 & 0 & 13 & \multirow{2}{*}{0.37} & \multirow{2}{*}{0.91} & 3 & 4 & 7 & \multirow{2}{*}{0.23} & \multirow{2}{*}{0.92} \\ \cline{2-3} \cline{8-11} \cline{14-16}
        ~ & Sky News & 4 & ~ & ~ & ~ & ~ & 17 & 2 & 1 & 20 & ~ & ~ & 1 & 5 & 6 & ~ & ~ \\ \hline
        \multirow{2}{*}{\textbf{Shopping}} & Amazon & 3 & \multirow{2}{*}{0} & \multirow{2}{*}{5} & \multirow{2}{*}{2} & \multirow{2}{*}{6} & 12 & 0 & 1 & 13 & \multirow{2}{*}{0.42} & \multirow{2}{*}{0.78} & 2 & 5 & 7 & \multirow{2}{*}{0.21} & \multirow{2}{*}{0.71} \\ \cline{2-3} \cline{8-11} \cline{14-16}
        ~ & Ebay & 4 & ~ & ~ & ~ & ~ & 21 & 0 & 2 & 23 & ~ & ~ & 3 & 4 & 7 & ~ & ~ \\ \hline
        \multirow{2}{*}{\textbf{Finance}} & Yahoo Finance & 4 & \multirow{2}{*}{1} & \multirow{2}{*}{5} & \multirow{2}{*}{2} & \multirow{2}{*}{6} & 15 & 0 & 1 & 16 & \multirow{2}{*}{0.40} & \multirow{2}{*}{0.81} & 3 & 3 & 6 & \multirow{2}{*}{0.25} & \multirow{2}{*}{0.88} \\ \cline{2-3} \cline{8-11} \cline{14-16}
        ~ & Stocklight & 3 & ~ & ~ & ~ & ~ & 14 & 0 & 2 & 16 & ~ & ~ & 1 & 1 & 2 & ~ & ~ \\ \hline
        \multirow{2}{*}{\textbf{Todo List}} & Microsoft Todo & 4 & \multirow{2}{*}{1} & \multirow{2}{*}{6} & \multirow{2}{*}{1} & \multirow{2}{*}{7} & 18 & 1 & 2 & 21 & \multirow{2}{*}{0.38} & \multirow{2}{*}{0.92} & 0 & 1 & 1 & \multirow{2}{*}{0.20} & \multirow{2}{*}{1.00} \\ \cline{2-3} \cline{8-11} \cline{14-16}
        ~ & Todoist & 3 & ~ & ~ & ~ & ~ & 12 & 1 & 2 & 15 & ~ & ~ & 2 & 2 & 4 & ~ & ~ \\ \hline
        \multirow{2}{*}{\textbf{Music}} & Spotify & 4 & \multirow{2}{*}{1} & \multirow{2}{*}{6} & \multirow{2}{*}{2} & \multirow{2}{*}{7} & 17 & 1 & 1 & 19 & \multirow{2}{*}{0.42} & \multirow{2}{*}{0.87} & 2 & 2 & 4 & \multirow{2}{*}{0.17} & \multirow{2}{*}{0.83} \\ \cline{2-3} \cline{8-11} \cline{14-16}
        ~ & Youtube Music & 4 & ~ & ~ & ~ & ~ & 17 & 1 & 1 & 19 & ~ & ~ & 1 & 1 & 2 & ~ & ~ \\ \hline
        \multirow{2}{*}{\textbf{Video}} & Ted & 4 & \multirow{2}{*}{0} & \multirow{2}{*}{4} & \multirow{2}{*}{2} & \multirow{2}{*}{6} & 22 & 0 & 1 & 23 & \multirow{2}{*}{0.41} & \multirow{2}{*}{0.76} & 1 & 1 & 2 & \multirow{2}{*}{0.22} & \multirow{2}{*}{0.67} \\ \cline{2-3} \cline{8-11} \cline{14-16}
        ~ & Youtube & 4 & ~ & ~ & ~ & ~ & 14 & 0 & 1 & 15 & ~ & ~ & 1 & 6 & 7 & ~ & ~ \\ \hline
        \multirow{2}{*}{\textbf{Map}} & Google Maps & 3 & \multirow{2}{*}{0} & \multirow{2}{*}{3} & \multirow{2}{*}{0} & \multirow{2}{*}{5} & 14 & 0 & 1 & 15 & \multirow{2}{*}{0.36} & \multirow{2}{*}{0.81} & 2 & 2 & 4 & \multirow{2}{*}{0.25} & \multirow{2}{*}{0.75} \\ \cline{2-3} \cline{8-11} \cline{14-16}
        ~ & Waze & 4 & ~ & ~ & ~ & ~ & 13 & 1 & 1 & 15 & ~ & ~ & 2 & 2 & 4 & ~ & ~ \\ \hline
        \multirow{2}{*}{\textbf{Photo}} & Instagram & 4 & \multirow{2}{*}{1} & \multirow{2}{*}{4} & \multirow{2}{*}{2} & \multirow{2}{*}{6} & 14 & 3 & 0 & 17 & \multirow{2}{*}{0.43} & \multirow{2}{*}{0.73} & 2 & 4 & 6 & \multirow{2}{*}{0.25} & \multirow{2}{*}{0.63} \\ \cline{2-3} \cline{8-11} \cline{14-16}
        ~ & Pinterest & 3 & ~ & ~ & ~ & ~ & 12 & 1 & 1 & 14 & ~ & ~ & 2 & 4 & 6 & ~ & ~ \\ \hline
        \multirow{2}{*}{\textbf{Social}} & Twitter & 4 & \multirow{2}{*}{1} & \multirow{2}{*}{5} & \multirow{2}{*}{1} & \multirow{2}{*}{6} & 18 & 0 & 2 & 20 & \multirow{2}{*}{0.42} & \multirow{2}{*}{0.89} & 3 & 3 & 6 & \multirow{2}{*}{0.20} & \multirow{2}{*}{1.00} \\ \cline{2-3} \cline{8-11} \cline{14-16}
        ~ & Linkedin & 3 & ~ & ~ & ~ & ~ & 15 & 0 & 1 & 16 & ~ & ~ & 1 & 3 & 4 & ~ & ~ \\ \hline
        \multirow{2}{*}{\textbf{Chat}} & Messenger & 4 & \multirow{2}{*}{0} & \multirow{2}{*}{6} & \multirow{2}{*}{1} & \multirow{2}{*}{8} & 21 & 1 & 2 & 24 & \multirow{2}{*}{0.36} & \multirow{2}{*}{0.91} & 0 & 2 & 2 & \multirow{2}{*}{0.33} & \multirow{2}{*}{1.00} \\ \cline{2-3} \cline{8-11} \cline{14-16}
        ~ & Teams & 4 & ~ & ~ & ~ & ~ & 18 & 3 & 3 & 24 & ~ & ~ & 0 & 1 & 1 & ~ & ~ \\ \hline
        \textbf{System} & Desktop & 3 & 1 & 1 & 1 & 2 & 17 & 3 & 0 & 20 & 0.44 & 0.85 & 2 & 2 & 4 & 0.25 & 0.75 \\ 
        \toprule
        
        \multicolumn{2}{c}{Total} & 107 & 10 (0.09) & \textbf{67 (0.63)} & 21 (0.20) & \textbf{95 (0.89)} & 448 & 22 & 37 & 507 & 0.40 & \textbf{0.86} & 49 & 83 & 132 & 0.23 & \textbf{0.85} \\ 

        \toprule
    \end{tabular}

    }
    \label{tab:eval_actions}
\vspace{-4mm}
\end{table*}

\subsubsection{NiCro's Performance}
NiCro takes the advantages of \textit{ResNet50} and \textit{Text Content} while matching non-text and text widgets, but also uses multi-modal attributes to further single out the most matched widget and GUIs for the \textit{Target Widget} and \textit{Target GUI}.
Although NiCro's matching approach performs significantly better overall, it produces incorrect matching in some situations in our experiment.
The most significant factor that affects the matching accuracy is the difference between the \textit{Source Device} and the \textit{Target Device}.
With the result shown in Figure~\ref{fig:eval_matching}, we observe the three most influential aspects of the devices: (1) screen size, (2) platform and (3) virtual or physical presence.
First, the difficulty of matching widgets increases when the screen size differences increases.
For example, the widget matching accuracy for widgets in \textit{D1}(8.0'') and \textit{D5}(3.7'') is noticeably lower than the accuracy for \textit{D3}(6'') and \textit{D4}(5'').
This is because the GUI varies with the device's screen size, which causes the change in GUI layout and widget's shape (e.g., size and aspect ratio).
Second, the widget's appearance and surrounding widgets may change dramatically on different platforms, which may affect NiCro's widget matching performance.
For instance, the performance is better when the widget in \textit{D6}(iOS) is matched to those in \textit{D7}(iOS) than to widgets on Android devices. 
Third, the virtual or physical presence of the compared devices may also impact the performance because the GUI image of physical devices is the photo taken in a natural environment where the environmental light takes some effects.
This impact can be observed from the experiment in which the widget matching accuracy for \textit{D2}(physical) and \textit{D3}(virtual) is always lower than the accuracy for \textit{D3}(virtual) and \textit{D4}(virtual) even though they are all Android devices and have similar screen size.
These effects can also be seen in the GUI matching which is based on the matched widgets.

\subsection{Action and Test Case Replay Performance}
This part of the evaluation focuses on the system's overall performance in replaying test cases and actions.
For each app, we applied 3 to 4 test cases composed of 3 to 8 GUI actions (widget-dependent or widget-independent) in every test case.
These test cases are all common usage scenarios that cover the general and specific functionalities of each app.
For example, "Change user name" for some apps that support user login, and "Search and enroll the most popular Python course" for the Education app.
This yields a total of 107 test cases containing 507 widget-dependent actions and 132 widget-independent actions.
In the process of the experiment, we randomly select one device as the \textit{Source Device} to record and then replay the actions on the rest of the 7 devices (\textit{Target Devices}).
Therefore, there are 639 (507 + 132) actions recorded and 4,473 (639 x 7) actions replayed.
We here use a strict replay success criterion: 
An action replay is successful if it is accurately reproduced on all 7 \textit{Target Devices}, and a test case is successfully replayed if all its actions are replayed correctly.

We observe three phenomena in these app usage scenarios.
(1) Widget-independent actions (i.e., swipe or scroll) are commonly involved in the test cases, in which 66 (61.7\%) of the 107 cases contain at least one widget-independent action when recorded.
This is because it is very common that the screen size is not enough to fill all the page content, which requires scrolling or swiping to move the rest content to the screen scope.
(2) The screen size variances across devices cause the recorded target widget on the \textit{Source Device} not shown on smaller \textit{Target Devices}.
It happens to 43 test cases (40.2\%) when they are replayed on some smaller devices, in which the screen needs to scroll to expose the target widget.
(3) The cross-platform implementation inconsistency changes the GUI layout and widget appearance in some apps.
78 test cases (72.9\%) experience inconsistency at various levels when replayed on different platforms and they may need more comprehensive information to match target widgets. 

With the observation, we compare NiCro with the three latest and closest works: MAPIT\cite{mapit},  LIRAT\cite{lirat} and RoScript~\cite{roscript}.
Table~\ref{tab:comparisons} shows the comparison.
It contrasts the three factors in the respective approaches (Cross-Platform; Cross-Device; Widget-Independent actions).
We also present the ratio of available test cases for each work in the \textbf{TC} column.
LIRAT and MAPIT do not support widget-independent actions, making them unavailable to handle 66 (61.7\%) test cases. 
Moreover, the intrusive technique MAPIT requires the target widget's metadata (e.g., accessibility id, text content etc.), while we find that MAPIT fails to obtain the information of 91 widgets (18\%) out of the 507 widgets in the widget-dependent actions on one or more devices, because of the absence or redundancy of specific widget attributes in the metadata.
Besides, the intrusive approach cannot acquire any content within an external component such as advertisement and WebView that is the primary element of some apps (e.g., web browsers).
This makes additional 12 test cases fail to run on MAPIT, which leaves only at most 29 test cases (27.1\%) available for it.
Although RoScript's approach solely considers the same UI on the same device, it supports both widget-dependent and widget-independent actions, and only requires the image as input, which enables it to run on all of the test cases in our experiment.
And thus, we use RoScript as our baseline to examine NiCro's overall performance.

Figure~\ref{tab:eval_actions} summarizes the experimental results.
Overall, NiCro achieves a promising performance for action replay, in which it accurately replays 86\% widget-dependent actions and 85\% widget-independent actions.
In contrast, RoScript only succeeds in 40\% and 23\% of these actions.
Regarding the test case replay, NiCro replays 63\% of them without any manual correction (\textbf{0-Correction}), where the entire test case replay fails even if only one of its action replay fails.
To investigate deeply, we examine the situation that allows one time of correction (\textbf{1-Correction}) for each test case and experiment NiCro and Rescript again, in which NiCro successfully replays 89\% cases and significantly outperforms its baseline (20\%).

Both of these two approaches' performances vary in different categories of apps.
In detail, NiCro performs better on apps containing more textual content than those containing more image elements.
It achieves 0.93 and 0.91 widget-dependent replay accuracy on Email and Chat apps while only reaching 0.76 and 0.73 on Video and Photo apps.
This is because it can better detect and match text widgets (as shown in Fig.~\ref{fig:eval_matching}), but for GUIs filled with complicated images or whose widgets overlap on images, its UIED-based widget detector cannot performs perfectly and hence causes cascading errors in widget matching.
Besides, the replay accuracy on Finance apps (0.81) and Map apps (0.81) is slightly lower for a similar reason, where they contain many diagrams and have GUI widgets overlapping on these diagrams, which fails the widget detection.
On the contrary, RoScrip's widget-dependent replay is even worse in textual apps (0.34 on Email apps and 0.36 on Chat apps) than in apps with more images (0.43 on Photo apps), because its Template Matching-based image matching approach cannot handle text-widget matching.
Overall, RoScript completes all widget-dependent action replay poorly also due to its incapable to properly work if the \textit{Target Widget} is invisible in the \textit{Target Device} due to the smaller screen, and is mostly incapable of matching widgets with different appearances across platforms.
On the other hand, NiCro's performance on widget-independent actions obviously surpasses RoScript, in which NiCro succeeds 112 cases out of 132 but RoScript only completes 30 cases.
This poor performance of RoScript is because it replays scroll and swipe simply with the same distance in the recording phase. 
However, such practice is inappropriate if the screen sizes of the recording device and the replay devices are inconsistent.
This confirms the effectiveness of NiCro's widget-independent action replay approach based on the GUI matching.

\subsection{Threats to Validity}
(1) The apps involved could be a threat.
To examine NiCro's generalization ability and counteract this, we use as many as 28 popular apps in 14 common categories plus the home page in the experiment.
However, there are more apps that would affect the experimental results, such as Game apps that are majorly composed of graphical elements, which the UIED-based detector would not handle perfectly~\cite{UIED-full-fse}.
(2) The tested devices could be another threat.
Although we intentionally select a diverse range of devices in terms of physical or virtual presence, different screen sizes and operating systems, it may still miss some particular devices that affect the result.
However, our evaluation confirms that NiCro can well perform the record and replay across physical and virtual devices with or without the same operating system.
(3) The robotic system's experimental environment could also be a threat.
We run NiCro's robotic system under a relatively stable environment with a steady light and fixed camera focus to acquire clear GUI photos.
The quality of the GUI photo taken by the camera may change under different circumstances and affect the widget detection and matching.
However, it is not difficult to adjust the physical setting to adapt to the new environment.
(4) The code of LIRAT and RoScript are manually re-implemented by ourselves.
In our communication with the authors of LIRAT and RoScript, they did not agree to release their source code and data for our comparative analysis purpose due to various restrictions, forcing us to re-implement their techniques. We tried our best and implemented them in a way that strictly followed the description in the original papers. 

\section{Related Work}

There is a range of intrusive testing frameworks for Android, such as Instrumentation~\cite{Instrumentation}, Robotium~\cite{Robotium} and Espresso~\cite{espresso}.
Monkeyrunner~\cite{Monkeyrunner} can replay the test script on several devices simultaneously.
Behrang et al~\cite{Behrang}. and Lin et al~\cite{lin}. attempt to apply static code analysis to extract the GUI models from Android apps to assist GUI testing, which is not suitable for closed-source systems, such as iOS.
For iOS, the mainstream automated testing tools include KIF~\cite{kif-framework} and UIAutomation~\cite{appium}.
Nevertheless, the intrusive tools for iOS are even harder to set up as it is not an open-source system, where the user has to pay a fee to apply for an Apple developer account and configure a set of software and hardware dependencies. Besides, all the above techniques do not support cross-platform testing.

TestMig~\cite{testmig} is the early attempt to migrate tests from iOS to Android.
However, this migration is a single-direction, which does not support replaying of the other direction, i.e., Android test scripts on iOS.
Appium~\cite{appium} uses Android Debug Bridge ~\cite{adb} and Web Driver Agent~\cite{webdriver} to communicate with Android and iOS devices through its Python-based test script.
MAPIT~\cite{mapit} is the latest work for cross-platform UI testing migration that enables the test migration from iOS to Android and the other way around. 
However, it still relies on the metadata to extract and match widgets and fails to tackle the cases where the metadata is not available or analyzable. 

Some works resort to image-based approaches for UI testing migration.
AppTestMigrator~\cite{AppTestMigrator} computes the similarity among GUI widgets and migrates test cases between apps with similar features.
Sikuli~\cite{sikuli} provides a visual test script that records the widget screenshot image and uses template matching algorithm~\cite{templatematching} to find the target widget on different devices while replaying.
LIRAT~\cite{lirat} is the latest approach to perform cross-platform automated UI testing based on GUI images.
In addition, some works utilize robot arms to physically interact with devices~\cite{robot1, rico2, roscript}.
Among them, RoScript~\cite{roscript} is the state-of-the-art for UI record-and-replay testing, which leverages image processing to recognise the human action in a recording video and uses template matching~\cite{templatematching} to locate the target widgets to replay.
However, these approaches are limited by their preliminary visual processing techniques and do not consider the GUI variance across devices with diverse screen sizes and operating systems.

\section{Conclusion and Future Work}
This paper presents a novel non-intrusive cross-device and cross-platform system named NiCro, to perform GUI action record and replay.
NiCro consists of a \textit{Device Farm} and a \textit{Robotic System} to support the interaction with virtual as well as physical devices.
It also utilizes a UI-specific widget detector and extracts comprehensive UI information to detect and match widgets and GUIs across various devices.
Our comprehensive evaluation demonstrates the promising of NiCro's as an early attempt of visual intelligence-based techniques toward more general non-intrusive software engineering tasks across platforms, architecture and devices.


\bibliographystyle{IEEEtran}
\bibliography{ref}

\end{document}